\documentclass[12pt]{article}
\usepackage{graphicx}
\usepackage{color}
\usepackage{cite}
\def \beq{\begin{equation}}
\def \eeq{\end{equation}}
\def\bea{\begin{eqnarray}}
\def\eea{\end{eqnarray}}

\begin{document}
\setcounter{footnote}{1}
\rightline{EFI 14-33}
\rightline{arXiv:1409.5813}
\vskip1.5cm

\centerline{\large \bf INTERPRETATION OF AN ``EDGE''}
\centerline{\large \bf IN PROTON-PROTON SCATTERING}
\bigskip

\centerline{
Jonathan L. Rosner\footnote{{\tt rosner@hep.uchicago.edu}}}
\medskip

\centerline{Enrico Fermi Institute and Department of Physics}
\centerline{\it University of Chicago, 5620 S. Ellis Avenue, Chicago, IL
60637, USA}
\bigskip
\strut

\begin{center}
ABSTRACT
\end{center}
\begin{quote}
A study of proton-proton collisions at very high energy has revealed a
 ``black disk,'' whose radius grows with the logarithm of the center-of-mass
energy, surrounded by an edge of approximately constant width 1 fm.  We
interpret this behavior as the maximum length of a QCD string connecting
the color triplet and antitriplet components of the proton, and propose
further tests of this explanation.
\end{quote}
\smallskip

\leftline{PACS codes: 11.80.Fv, 13.85.Lg, 13.85.Dz, 13.75.Cs}
\bigskip

It has been known for over fifty years that the total proton-proton cross
section cannot grow with energy faster than $\ln^2 s$, where $s$ is the
square of the center-of-mass energy \cite{Froissart:1961ux, Martin:1962rt}.
Recently fits to $pp$ and $\bar p p$ cross sections up to Tevatron energies
have indicated a structure consisting of a black disk whose radius grows
as $\ln \sqrt{s}$, surrounded by an edge of approximately constant thickness
of $\sim 1$ fm \cite{Block:2014lna}.  These fits describe data both from the
CERN Large Hadron Collider (LHC) and from cosmic ray interactions up to
$\sqrt{s} = 80$ TeV \cite{Block:2005ka,Block:2011vz,Block:2012nj}.

The purpose of this note is to interpret the energy-independent edge as the
extent to which a color-triplet constituent of the proton can separate from the
remainder (which will be an antitriplet) before the QCD string connecting them
breaks.  This provides a fuzzy ``edge'' to the proton, or, for that matter, to
any hadron which can dissociate into a 3-$3^*$ virtual pair.  We begin by
recalling the arguments for a universal ``string-breaking'' length, apply them
to proton-proton scattering, and propose further tests of this prediction of
universal behavior.

\begin{figure}
\begin{center}
\includegraphics[width=0.8\textwidth]{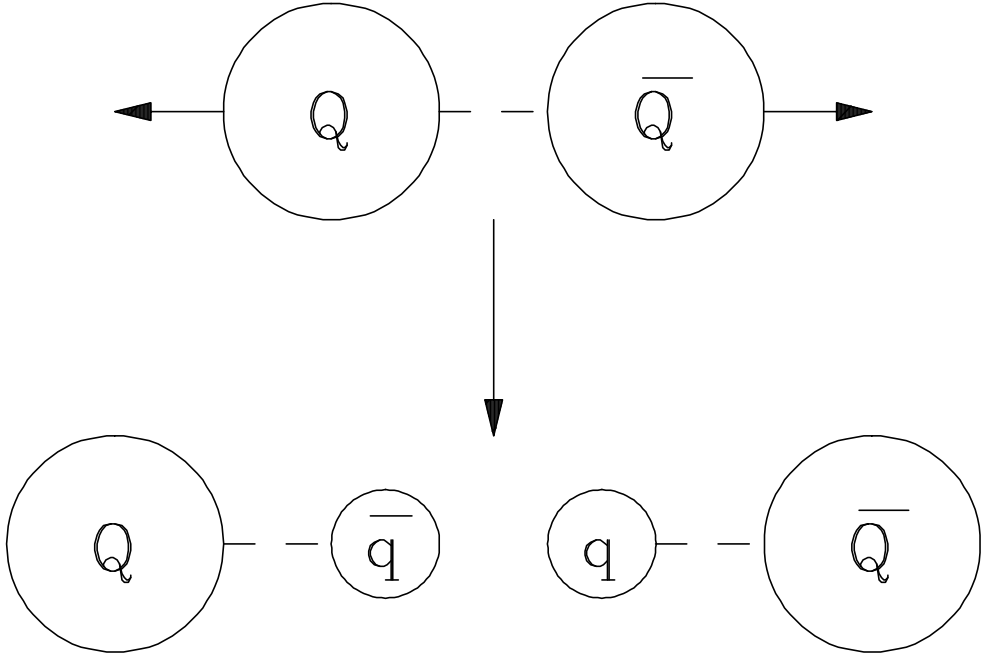}
\end{center}
\caption{Top:  Heavy quarks $Q$ and $\bar Q$ connected by a QCD string
(dashed line).  When pulled sufficiently apart (bottom) it is energetically
favorable for $Q$ to become connected to a light antiquark $\bar q$ and
$\bar Q$ to become connected to a light quark $q$ by separate strings.
\label{fig:break}}
\end{figure}

A heavy quark $Q$ and a heavy antiquark $\bar Q$ may be viewed as connected
by a color-triplet QCD string, as shown in Fig.\ \ref{fig:break}.  When
$Q$ and $\bar Q$ are pulled apart by a sufficient distance, the QCD string
connecting them will contain sufficient energy to permit the production of
a light quark-antiquark pair $\bar q q$, with $\bar q$ connected to $Q$ and
$q$ connected to $\bar Q$ by separate QCD strings.

The flavor-independence of the $Q \bar Q$ potential was demonstrated
some time ago for charmonium and bottomonium systems \cite{Quigg:1981bj,
Buchmuller:1980su}.  Additional recent supporting evidence has come from
the observation of an excited $\bar b c$ system separated from the
ground state $B_c$ by an amount between the (roughly equal) 2S--1S
spacings of charmonium and bottomonium \cite{Aad:2014laa}.  Thus it
suffices to base an argument on the bottomonium potential, which contains
the most levels below flavor threshold and thus for which the most detailed
information is available \cite{Rosner:1996xz}.  Using an approximate
simple form for the interquark potential \cite{Quigg:1977dd}, and taking note
of the bottomonium flavor threshold of 10.58 GeV, it was found that the
$b$--$\bar b$ separation for which the string between them breaks lay between
1.4 and 1.5 fm.  At that time unquenched lattice QCD was just entering its
infancy, so no corresponding lattice result was available.

The situation has now changed.  A detailed QCD lattice calculation with two
light flavors \cite{Bali:2005fu} estimates a threshold separation
$r_c = (1.13\pm0.10\pm0.10)$ fm, including all systematic errors.  One expects
the addition of the strange quark flavor to have little effect as
$B_s$--$\bar B_s$ threshold lies considerably higher than the $B$--$\bar B$
threshold of 10.56 GeV.

How does this relate to the thickness of the edge seen in Ref.\
\cite{Block:2014lna}?  One imagines any hadron to consist of quarks,
antiquarks, and gluons, each of which can be imagined to be on a ``leash''
connecting them to the rest of the hadron.  Color triplets will be connected
to antitriplet remnants; color antitriplets will be connected to triplet
remnants; and gluons will be connected to color octet remnants.  We shall
assume that the color octet string connecting a gluon with the rest of the
proton has greater tension and thus plays less of a role in determining the
proton's outer edge.  Then the maximum radius of the proton, no matter at what
energy, will fluctuate from a value $R$ to a value $R + \Delta$ where $\Delta$
corresponds to the maximum length of a color-triplet QCD string.

The proton-proton scattering amplitude at high energy is predominantly
imaginary, $a = (\eta - 1)/(2i)$, where $\eta = 0$ corresponds to maximum
absorption and $\eta = 1$ corresponds to no scattering.  The total and elastic
cross sections may be expressed as integrals over an impact parameter $b$ as
\beq
\sigma_{\rm tot} = 8 \pi \int_0^\infty (1-\eta)~b~db/2~~,~~~
\sigma_{\rm el} = 8 \pi \int_0^\infty (1 - \eta)^2~b~db/4~.
\eeq
A simplified model of the behavior noted in Ref.\ \cite{Block:2014lna} may
be constructed for an imaginary scattering amplitude with $\eta = 0$ for an
impact parameter $b < R$, $\eta = (b-R)/\Delta$ for $R \le b \le R+\Delta$,
and $\eta = 1$ for $b > R+\Delta$.  Then for $\Delta \ll R$,
\beq
\sigma_{\rm tot} - 2 \sigma_{\rm el} = 4 \pi \int_{R}^{R+\Delta}b~db~
\eta~(1-\eta) \simeq 4 \pi R \Delta \int_0^1 \eta~(1-\eta)~d \eta
= 2 \pi R \Delta / 3~.
\eeq
This quantity is denoted $\pi R t$ in Ref.\ \cite{Block:2014lna}, where $t
\simeq 1.1$ fm is interpreted as the thickness of the edge.  Thus, in
our language, $\Delta = 3t/2 \simeq 1.6$ fm, close to the original estimate
of Ref.\ \cite{Rosner:1996xz} for the string-breaking distance. 

The question is asked in Ref.\ \cite{Block:2014lna}:  How universal is this
behavior?  In order to test it, one would have to investigate other
systems besides proton-proton collisions.  But proton-proton collisions
can be used to generate effective pion beams, as in the reactions
$pp \to n \pi^+ p$ \cite{Ming:1969pf} and $pp \to \Delta^{++} X$
\cite{Ming:1969pf,Lockman:1976rj}, where $X$ has the quantum numbers of $\pi^-
p$.  We thus advocate studying the latter reaction at the LHC with the
decay $\Delta^{++} \to \pi^+ p$ detected at low momentum transfer from
the initial proton, in such a way that the effect of the pion pole can
be isolated.  As the pion also can be separated into a color triplet and
a color antitriplet, we predict that pion--nucleon scattering will
exhibit the same edge behavior as seen in proton-proton scattering.

Another possibility for studying very-high-energy elastic cross sections
would be the electroproduction on protons at low $Q^2$ of the vector mesons
$(\rho,\omega,\phi)$, which would occur in the ratio 9:1:2 by virtue of
the corresponding quark charges.  However, it would be more challenging
to separate the total cross sections of individual vector mesons on protons
from such data.
 
I thank Martin Block, Andreas Kronfeld, Wolfgang Ochs, and Leo Stodolsky for
discussions; the W. and E. Heraeus Foundation for partial support during a
workshop at Oberw\"olz, Austria; the Max-Planck-Insit\"ut f\"ur Physik for
hospitality during part of this investigation, and the Physics Department of
the University of Chicago for partial travel support.  This work was
supported in part by the U. S. Department of Energy under Grant No.\
DE-FG02-13ER41598.

\end{document}